\documentclass[floatfix,superscriptaddress,a4paper,
               showpacs,showkeys,nofootinbib,preprint]{revtex4}
\usepackage{epsfig}
\usepackage{latexsym}
\usepackage{xspace}
\usepackage{xcolor}
\usepackage{hyperref}
\usepackage[latin2]{inputenc}
\usepackage{indentfirst}
\usepackage{enumerate}

\usepackage[normalem]{ulem}
\newcommand{\stkout}[1]{\ifmmode\text{\sout{\ensuremath{#1}}}\else\sout{#1}\fi}

\usepackage{amsmath}
\usepackage{amssymb}
\usepackage[english]{babel}
\usepackage{url}
\newcommand{\eq}[1]{\begin{align} #1 \end{align}}

\begin{document}

\title{
Multicomponent van der Waals equation of state:
\\
Applications in nuclear and hadronic physics
}
\author{Volodymyr Vovchenko}
\affiliation{
Institut f\"ur Theoretische Physik,
Goethe Universit\"at Frankfurt, D-60438 Frankfurt am Main, Germany}
\affiliation{
Frankfurt Institute for Advanced Studies, Goethe Universit\"at Frankfurt,
D-60438 Frankfurt am Main, Germany}
\affiliation{Department of Physics, Taras Shevchenko National University of Kiev, 03022 Kiev, Ukraine}
\author{Anton Motornenko}
\affiliation{Department of Physics, Taras Shevchenko National University of Kiev, 03022 Kiev, Ukraine}
\affiliation{Department of Physics, University of Oslo, 0313  Oslo, Norway}
\affiliation{
Frankfurt Institute for Advanced Studies, Goethe Universit\"at Frankfurt,
D-60438 Frankfurt am Main, Germany}
\author{Paolo Alba}
\affiliation{
Frankfurt Institute for Advanced Studies, Goethe Universit\"at Frankfurt,
D-60438 Frankfurt am Main, Germany}
\author{Mark~I.~Gorenstein}
\affiliation{
Bogolyubov Institute for Theoretical Physics, 03680 Kiev, Ukraine}
\affiliation{
Frankfurt Institute for Advanced Studies, Goethe Universit\"at Frankfurt,
D-60438 Frankfurt am Main, Germany}
\author{Leonid M. Satarov}
\affiliation{
Frankfurt Institute for Advanced Studies, Goethe Universit\"at Frankfurt,
D-60438 Frankfurt am Main, Germany}
\affiliation{
National Research Center ''Kurchatov Institute'' 123182 Moscow, Russia}
\author{Horst Stoecker}
\affiliation{
Institut f\"ur Theoretische Physik,
Goethe Universit\"at Frankfurt, D-60438 Frankfurt am Main, Germany}
\affiliation{
Frankfurt Institute for Advanced Studies, Goethe Universit\"at Frankfurt,
D-60438 Frankfurt am Main, Germany}
\affiliation{
GSI Helmholtzzentrum f\"ur Schwerionenforschung GmbH, D-64291 Darmstadt, Germany}

\date{\today}

\begin{abstract}
A generalization of the quantum van der Waals equation of state
for a multi-component system in the grand canonical ensemble is proposed.
The model includes quantum statistical effects and allows to specify the
parameters characterizing repulsive and attractive forces
for each pair of particle species.
The model can be straightforwardly applied to the description of asymmetric nuclear
matter and also for mixtures of interacting nucleons and nuclei.
Applications of the model to the equation of state of an interacting hadron resonance gas are discussed.
\end{abstract}



\maketitle

\section{Introduction}
The equation of state (EoS) of  hot, dense strongly interacting matter
is in the focus of experimental and theoretical investigations of high-energy heavy-ion collisions.
Thermal models were constructed~\cite{Mekjian:1977ei,Gosset:1988na,Mekjian:1978us,Stoecker:1981za,Csernai:1986qf,Hahn:1987tz,Hahn:1986mb,Cleymans:1992zc,BraunMunzinger:1996mq,Becattini:2000jw}
to describe yields of secondary hadrons and nuclear fragments from such collisions.
These  models  assume that the emitted particles stem from a statistically
equilibrated system. The temperature $T$ and baryon chemical
potential $\mu_B$ of the emitting source are obtained by fitting the observed yields of stable hadrons.
In most cases, the ideal hadron resonance gas (I-HRG) model has been used.
A surprisingly good description of many experimental hadron yield data
from heavy-ion collisions have been achieved within this
simple approach for a broad range of collision energies
(see, e.g., \cite{Braun-Munzinger:2015hba} and references therein). The $T$-$\mu_B$ values fitted
in the I-HRG model show, however, that hadron densities are rather large at the chemical
freeze-out stage of the reaction. 
Therefore,
one can expect 
residual interactions,
leading to
significant deviations from the ideal gas picture.

Phenomenological models of the phase structure of nuclear matter also show~\cite{Rischke:1991ke,Satarov:2009zx}
that a realistic phase diagram cannot be obtained without an explicit account
of the hadronic interactions.
Extensions of the ideal gas picture have been
discussed mostly within the excluded volume HRG model, in which the
effects of various hadron's repulsions at
short distances have been introduced (see e.g., \cite{Satarov:2016peb} and references
therein).

The presence of both, repulsive and attractive interactions between nucleons, is evident from the existence of stable nuclei. 
These interactions should also be taken into account
to describe the simple binding properties of nuclear matter,
and the multifragmentation observed
at intermediate energy nucleus-nucleus
collisions. 
Recently, nuclear matter 
has been modeled
as van der Waals (vdW) fluid of interacting
nucleons~\cite{Vovchenko:2015vxa}. 
The vdW parameters $a$ and $b$ describe, respectively, attractive
and repulsive vdW interactions. $a$ and $b$ are fixed by fitting the nucleon number density and binding energy
per nucleon at $T=0$.
Attractive and repulsive interactions between nucleons were also discussed
within the mean-field theory~(see~\cite{Anchishkin:2014hfa} and references therein).
vdW interactions
between baryons, and between antibaryons, respectively, were considered within a quantum vdW-HRG model in Ref.~\cite{Vovchenko:2016rkn}.

This paper presents general formulation of the quantum van der Waals (QvdW) model, with different repulsive and attractive interactions in the multi-component system of different particle species. 
Classical multi-component vdW description is done for the pressure as a function of the temperature $T$ and the particle number densities $n_i$ ($i=1,\ldots,h$) in the canonical ensemble (CE): $p=p(T,n_1,\ldots,n_h)$.
To proceed further, the free energy $F$ is reconstructed.
$F$ is the thermodynamical potential in the CE. 
At this stage, we introduce quantum statistics. 
Note that a quantum vdW model was previously formulated for a single constituent type only~\cite{Vovchenko:2016rkn}.

Next, the model formulation is transformed to the grand canonical ensemble (GCE).
Ref.~\cite{Vovchenko:2015xja} shows that this transformation is important for several reasons.
Note that the chemical potentials -- and \emph{not} the particle densities --
and the temperature are the natural variables of the pressure function
$p=p(T,\mu_1,\ldots,\mu_h)$. 

Sec.~\ref{CE} presents the multi-component vdW
model in the CE,
and 
introduces the effects of quantum statistics.
Sec.~\ref{GCE} presents the GCE formulation.
Sec.~\ref{Appl} discusses the applications of the developed formalism.
Many applications of the multi-component QvdW equation are in sight: A first example is
a model of nuclear matter that includes a mixture of interacting protons, neutrons,
and nuclei. A second example is the multi-component QvdW HRG.
Sec. \ref{Sum} summarizes the paper.

\section{Canonical ensemble formulation}\label{CE}

The classical multi-component system with vdW interactions
is defined within the
CE in terms of the following pressure function:
\eq{\label{vdw-p}
p(T,n_1,\ldots,n_h)~ =~ \sum_i \frac{T\,n_i}{1-\sum_j \tilde{b}_{ji}\,n_j} ~-~ \sum_{i,j} a_{ij} \, n_i \, n_j~,
}
where $n_i$ is the particle density for the $i$th species ($i=1,\ldots, h$) and $T$ is the system's temperature.
The parameters $a_{ij}$ and $\tilde{b}_{ji}$ in Eq.~(\ref{vdw-p}) yield the  attractive and repulsive vdW
interactions, respectively. 
The repulsive vdW interactions amount to the
excluded volume (EV) correction.
For particles with the classical hard-core interaction one has~\cite{Gorenstein:1999ce}
\eq{ \label{bij}
\tilde{b}_{ij} = 2 b_{ii} b_{ij}/(b_{ii}+b_{jj})~,
}
where $b_{ij} = 2\pi \, (r_i + r_j)^3/3$ is the symmetric matrix of the 2nd order virial coefficients, $r_i$ being the hard-core radius for the $i$th particle species.
In a more general case, $\tilde{b}_{ij}$ should be regarded as phenomenological parameters, characterizing the strength of the repulsive interactions between different pairs of particle species.

The free energy $F(T,V,\{N_i\})$ in the CE reads~\cite{Satarov:2016peb}:
\eq{\label{F}
F(T,V,\{N_i\})~ = ~\sum_i F^{\rm id}_i (T,V - \sum_j \tilde{b}_{ji} N_j, N_i) ~-~ \sum_{i,j} a_{ij} \frac{N_i \, N_j}{V}~.
}
The function  $F^{\rm id}_i (T,V, N_i)$ is the free energy of the classical (Boltzmann) ideal
gas for species $i$ ($\hbar = c =1$):
\eq{\label{F-id}
F^{\rm id}_i (T,V, N_i)~=~-~N_i\,T\left[1~+~\ln \frac{g_i\,V\,m_i^2\,T\,K_2(m_i/T)}{2\pi^2\,N_i}\right]~.
}
Here $m_i$ is the mass of particle species $i$,  $g_i$ is the particle  degeneracy factor (i.e., the number of internal states),
and $K_2$ is the modified Bessel function.
Note that the free energy $F$ is a genuine thermodynamic potential in the CE. $F$ gives the complete
information about the statistical system considered.
The partial derivative of $F$ (\ref{F}) over the system volume, $p=-(\partial F/\partial V)_{T,\{N_i\}}$,
yields the vdW pressure (\ref{vdw-p}). Note that free energy (\ref{F}) depends on the parameters $\{m_i\}$ and
$\{g_i\}$, but these parameters are absent in the vdW pressure (\ref{vdw-p}) in the CE.

Expressions (\ref{vdw-p}) and (\ref{F}) do not include the effects of quantum statistics.
The classical free energies of an ideal gas $F_i^{\rm id}$ in Eq.~(\ref{F-id})
are replaced by the ideal {\it quantum} gas expressions (Fermi-Dirac or Bose-Einstein) in order to include these effects.
This procedure satisfies the following consistency requirements: It leads to a mixture of the ideal quantum gases if
all $\tilde{b}_{ij}=0$ and $a_{ij}=0$. It gives the correct limiting classical expressions (\ref{vdw-p}) in the regions
of the thermodynamic parameters where quantum statistics can be neglected. Finally,
the entropy obtained for the QvdW expressions is positive definite, $S \geq 0$, and it 
respects Nernst's theorem, $S \to 0$ at $T \to 0$.
Quantum statistics of the one-component vdW fluid has been considered in Refs.~\cite{Vovchenko:2015vxa,Poberezhnyuk:2015dba,Redlich:2016dpb}.

Using this quantum statistical free energy, one can calculate all other thermodynamic functions in the CE: pressure $p$,
total entropy $S$ and energy $E$ of the system, as well as the $i$th chemical potential $\mu_i$:
\eq{\label{p}
p(T,\{n_k\})~& \equiv~-\,\left(\frac{\partial F}{\partial V}\right)_{T,\{N_j\}}
~=~ \sum_i p^{\rm id}_i \left(T, \frac{n_i}{f_i}\right) - \sum_{i,j} a_{ij} \, n_i \, n_j~,\\
S(T,V,\{N_k\})~&\equiv ~-\,\left(\frac{\partial F}{\partial T}\right)_{V,\{N_j\}}
~ =~ V \sum_i f_i\, s^{\rm id}_i \left(T, \frac{n_i}{f_i}\right)~,\label{S}\\
E(T,V,\{N_k\}) ~& \equiv ~ F+TS
~=~ V \sum_i f_i \, \varepsilon^{\rm id}_i \left(T, \frac{n_i}{f_i}\right)- V \sum_{i,j} a_{ij} \,n_i \, n_j~,
\label{E}\\
\mu_i(T,\{n_k\})~&\equiv~ \left(\frac{\partial F}{\partial N_i}\right)_{T,V,\{N_{j \neq i}\}}~ =~ \mu^{\rm id}_i \left(T, \frac{n_i}{f_i}\right)
~+~
\sum_j \tilde b_{ij} \, p^{\rm id}_j \left(T, \frac{n_j}{f_j}\right)
~-~
\sum_j (a_{ij} + a_{ji}) \, n_j~.
\label{muk}
}
Here $f_i \equiv 1-\sum_j\tilde{b}_{ji}n_j$ quantifies the fraction of the total volume which is available for particles at the given density value, and
$p^{\rm id}_i$,  $s^{\rm id}_i$, $\varepsilon^{\rm id}_i$, and  $\mu^{\rm id}_i$ are the ideal quantum
gas expressions for the CE pressure, entropy density, energy density, and
chemical potential for $i$th particle species, respectively, as functions of temperature  and particle number density.

\section{Grand canonical ensemble formulation}
\label{GCE}
In the GCE, the thermodynamic variables are the temperature $T$ and the set of chemical potentials $\{ \mu_k \}$.
The particle densities $\{ n_k \}$, on the other hand, are not the independent variables anymore in the GCE. Instead, they become functions of the $T$ and $\{ \mu_k \}$.
The pressure function $p(T, \{ \mu_k \})$ defines all thermodynamic properties of the system in this ensemble.

Both the CE and the GCE are equivalent for describing the thermodynamic properties in the thermodynamic limit, $V \to \infty$.
This fact can be used to transform the multi-component QvdW model from the CE to the GCE: it is sufficient only to derive the function $p(T, \{ \mu_k \})$ from Eqs.~\eqref{p}-\eqref{muk}.

First, let us introduce the notations
\eq{\label{*}
p^*_i \equiv p^{\rm id}_i \left(T, \frac{n_i}{f_i}\right), \quad
n^*_i \equiv n^{\rm id}_i \left(T, \frac{n_i}{f_i}\right),  \quad
s^*_i \equiv s^{\rm id}_i \left(T, \frac{n_i}{f_i}\right),
}
and
\eq{\label{*2}
\mu^*_i \equiv \mu^{\rm id}_i \left(T, \frac{n_i}{f_i}\right).
}
Here all the ideal gas functions correspond to the CE.
Equation~(\ref{*2}) can be inverted to yield:
\eq{\label{ni1}
\frac{n_i}{f_i} = n^{\rm id}_i (T, \mu^*_i)~,~~~ \quad i=1,\ldots,h~,
}
where $n^{\rm id}_i (T, \mu^*_i)$ is now the GCE ideal gas density at temperature $T$ and chemical potential $\mu^*_i$.
Using~\eqref{ni1} one can rewrite Eqs.~\eqref{*} as
\begin{equation}
p^*_i = p^{\rm id}_i \left(T, \mu^*_i\right)~, \quad
n^*_i = n^{\rm id}_i \left(T, \mu^*_i\right)~,  \quad
s^*_i = s^{\rm id}_i \left(T, \mu^*_i\right)~, \label{**}
\end{equation}
where all the ideal gas functions in~\eqref{**} correspond to the GCE.
Therefore, the GCE pressure reads
\eq{\label{pressure}
p(T, \{ \mu_i \}) ~=~ \sum_i p^*_i ~-~ \sum_{i,j} a_{ij} \, n_i \, n_j~.
}

Equations~\eqref{ni1} can be rewritten as the system of linear equations for particle densities $n_i$:
\eq{\label{ni}
\sum_j (\delta_{ij} ~+~ \tilde{b}_{ji} \, n^*_i) \, n_j ~=~ n^*_i~, ~~\quad i = 1,\ldots,h~.
}

If the $\mu^*_i$, at given $T$ and $\mu_i$, are known, then all other quantities, in particular,
the system pressure (\ref{pressure}), can be calculated as well. Indeed, the calculations
of $p^*_i$ and $n^*_i$ are straightforward, while $n_i$ can be recovered by solving the system of linear equations~\eqref{ni}.
Finally, the pressure is obtained by substituting $p^*_i$ and $n_i$ into~\eqref{pressure}.
All other thermodynamic functions in the GCE are obtained from the pressure function $p(T,\{\mu_i\})$ using standard thermodynamic relations.

Using the above notations, one can rewrite~\eqref{muk} as
\eq{\label{mu*}
\mu^*_i + \sum_j \tilde b_{ij} \, p^*_j - \sum_j (a_{ij} + a_{ji}) \, n_j = \mu_i~, \quad i = 1,\ldots,h~.
}
Solution to this system of transcendental equations determines $\{\mu_i^*\}$ at given $T$ and $\{\mu_i\}$.
In general, this solution should be obtained numerically.
Once the effective chemical potentials $\{\mu_i^*\}$ are specified, all other quantities can be calculated directly.
If multiple solutions of equations (\ref{mu*}) are found, the solution with the largest pressure 
is the physical result
according to the Gibbs criterion.

The GCE entropy and energy densities are given as
\eq{
s(T, \{ \mu_i \})~& \equiv~\left(\frac{\partial p}{\partial T}\right)_{\{\mu_j\}}~ =~ \sum_i f_i \, s^*_i~,\\
\varepsilon(T, \{ \mu_i \})~& \equiv~\left(\frac{\partial p}{\partial T}\right)_{\{\mu_j\}}+\sum_i \mu_i\left(\frac{\partial p}{\partial \mu_i}\right)_{T,\{\mu_{j\neq i}\}}-p~
 = \sum_i f_i\, \varepsilon^*_i- \sum_{i,j} a_{ij} \,n_i \, n_j~.
}
Here $\varepsilon_i^*=\varepsilon_i^{\rm id}(T,\mu_i^*)$.

\section{Applications}
\label{Appl}

The new multi-component QvdW
formalism presented here is applied to the following
three examples:

\subsection{Asymmetric nuclear matter as a QvdW mixture of protons and neutrons}

Nuclear matter is a hypothetical infinite system of interacting protons and neutrons which is approximately realized in nature in the interior of massive nuclei and in neutron stars.
The thermodynamic equilibrium in such a system can be specified by the temperature, $T$, and proton and neutron densities, $n_p$ and $n_n$.
Nucleon-nucleon interactions  exhibit repulsion at short distances and attraction at intermediate ones. Hence, it makes sense to model such a system by a QvdW equation.
The multi-component QvdW equation for protons and neutrons then reads
\eq{\label{eq:P-ASNM}
p(T,n_p,n_n) = p^{\rm id}_p \left(T, \frac{n_p}{f_p}\right) + p^{\rm id}_n \left(T, \frac{n_n}{f_n}\right) - a_{pp} \, n_p^2 - a_{pn} \, n_p \, n_n - a_{np} \, n_n \, n_p - a_{nn} \, n_n^2,
}
where $f_p = 1 - \tilde{b}_{pp} \, n_p - \tilde{b}_{np} \, n_n$ and $f_n = 1 - \tilde{b}_{pn} \, n_p - \tilde{b}_{nn} \, n_n$.
Isospin symmetry yields $m_p = m_n \simeq 938$~MeV/$c^2$, $\tilde{b}_{pp} = \tilde{b}_{nn}$,
$\tilde{b}_{pn} = \tilde{b}_{np}$, $a_{pp} = a_{nn}$, and $a_{pn} = a_{np}$.
This model has only four interactions parameters: $\tilde{b}_{pp}$, $\tilde{b}_{pn}$, $a_{pp}$, and $a_{pn}$. These correspond to isospin-dependent nucleon-nucleon interactions.

The asymmetry of the nuclear matter is characterized by the proton fraction $y = n_p / (n_p + n_n)$, which takes values between 0 and 1, and the total nucleon density $n_N = n_p + n_n$.
The value $y = 1/2$ corresponds to the symmetric nuclear matter, i.e. the numbers of protons and neutrons in the system are equal. In this case, Eq.~\eqref{eq:P-ASNM} reduces to a single-component QvdW equation for nucleons,
\eq{\label{eq:P-SNM}
p(T,n_N,y=1/2) = p^{\rm id}_N \left(T, \frac{n_N}{1 - b_{NN} \, n_N}\right) - a_{NN} \, n_N^2.
}
Here the nucleon ideal gas pressure $p^{\rm id}_N$ contains the spin-isospin nucleon degeneracy factor $g_N = 4$, and
\eq{
a_{NN} = \frac{a_{pp} + a_{pn}}{2} \quad \textrm{and} \quad \tilde{b}_{NN} = \frac{\tilde{b}_{pp} + \tilde{b}_{pn}}{2}.
}
Equation~\eqref{eq:P-SNM} coincides with the model used in Ref.~\cite{Vovchenko:2015vxa}. The parameters $a_{NN}$ and $\tilde{b}_{NN}$ are fixed by reproducing the binding energy $\varepsilon / n_N - m_N = -16$~MeV at saturation density $n_0 = 0.16$~fm$^{-3}$. The following parameter values are obtained:
$a_{NN} \simeq 329$ MeV fm$^3$ and $\tilde{b}_{NN} \simeq 3.42$~fm$^{3}$.

In the present work these $a_{NN}$ and $\tilde{b}_{NN}$ values are used, while the $a_{pn} / a_{pp}$ and $\tilde{b}_{pn} / \tilde{b}_{pp}$ ratios are taken as free parameters.
Variations in these ratios correspond to different scenarios for the isospin dependence of the nucleon-nucleon potential.

Cold nuclear matter with $T = 0$ is considered.
The nuclear symmetry energy $S(n)$ is sensitive to the isospin part of the nucleon-nucleon interactions. 
It characterizes the dependence of the energy per nucleon $E/A \equiv \varepsilon / n_N - m_N$ on the proton fraction $y$. The widely used parabolic approximation represents $E/A$ as
\eq{\label{eq:EAsym}
E/A \, (n, y) \approx E/A \, (n, y=1/2) + 4 \, S(n) \, (y-1/2)^2.
}

\begin{figure}[t]
\centering
\includegraphics[width=0.75\textwidth]{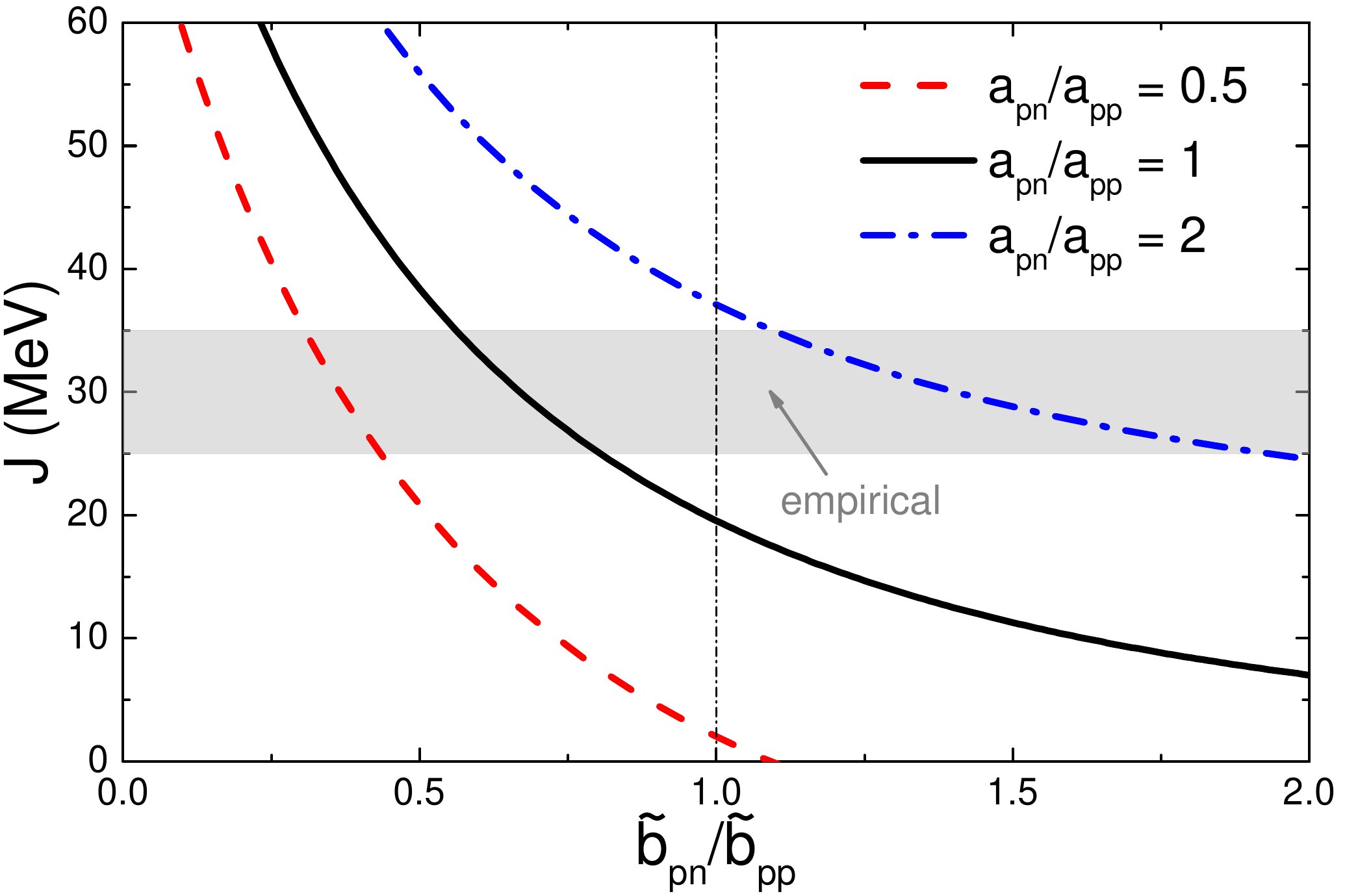}
\caption[]{
Dependence of the symmetry energy $J$ on the ratio $\tilde{b}_{pn} / \tilde{b}_{pp}$ of the repulsive vdW parameters for three different values of the ratio $a_{pn} / a_{pp}$ of the attractive vdW parameters. 
Variations in these ratios probe the isospin dependence of the nucleon-nucleon potential.
}\label{fig:NMJ}
\end{figure}

The symmetry energy at saturation density, defined as
\eq{\label{eq:J}
J \equiv S(n_0) = \frac{1}{8} \left. \frac{\partial^2 (E/A)}{\partial y^2}\right|_{n_N = n_0,\,y = 1/2},
}
corresponds roughly to the difference of the energy per nucleon at $n_N = n_0 \simeq 0.16$~fm$^{-3}$ for pure neutron matter~($y=0$) and symmetric nuclear matter~($y=1/2$).

The dependence of $J$ on different values of the $a_{pn} / a_{pp}$ and $\tilde{b}_{pn} / \tilde{b}_{pp}$ ratios is shown in Fig.~\ref{fig:NMJ}. 
The empirical range $J \simeq 25-35$~MeV~\cite{Dutra:2014qga} is depicted by the shaded area.
In the fully symmetric scenario, i.e. for $a_{pp} = a_{pn} = a_{NN}$ and $\tilde{b}_{pp} = \tilde{b}_{pn} = \tilde{b}_{NN}$, the value of the symmetry energy $J \simeq 18$~MeV underestimates significantly the empirical estimates.
In this case the total symmetry energy value is attributed solely to the decrease of the spin-isospin degeneracy factor, from 4 in symmetric nuclear matter to 2 in pure neutron matter.
This mechanism is not sufficient to describe the empirical data.
On the other hand, either an increase in $a_{pn}$ or a reduction in $\tilde{b}_{pn}$ improves the agreement with the data.

More stringent restrictions on the values of parameters $a_{pn}$ and $\tilde{b}_{pn}$ can be obtained by analyzing additional observables. 
These may include the density dependence of the symmetry energy, or the higher order terms of expansion~\eqref{eq:EAsym} of the $E/A$ in terms of the proton fraction $y$. 

\subsection{Mixture of interacting nucleons and $\alpha$ particles}

Light nuclei ought to be included in models of nuclear matter. 
A simple example
is a mixture of nucleons and $\alpha$ particles in symmetric nuclear matter, i.e. $n_p = n_n$.
The baryon number $B=N_N+4N_{\alpha}$
is conserved in this system, but
the numbers of nucleons, $N_N$,  and  $\alpha$'s, $N_\alpha$, are not conserved separately.
The chemically equilibrated $N-\alpha$ mixture has one independent
baryonic chemical potential $\mu$ which regulates the baryonic density $n\equiv n_N+4n_\alpha$.
The chemical potentials of nucleons and alphas are $\mu_N=\mu$ and $\mu_\alpha=4\mu$,
the $\alpha$ binding energy is contained in the $\alpha$ mass.
The pressure of the system is~(\ref{vdw-p}):
\eq{\label{p-Nalpha}
p ~=~ p^*_N~ +~ p^*_\alpha ~-~ a_{NN}\,n_N^2~-~a_{N\alpha}\,n_N n_\alpha~-~a_{\alpha N}\,n_\alpha n_N ~-~a_{\alpha\alpha}\,n_\alpha^2~.
}
Here
\eq{\label{pN-id}
p_N^* & \equiv~p_N^{\rm id}(T,\mu_N^*)=\frac{g_N}{6\pi^2}\int_0^\infty \frac{dk\,k^4}{\sqrt{m_N^2+k^2}}\,
\left[\exp\left(\frac{\sqrt{m_N^2+k^2}-\mu_N^*}{T}\right)~+~1\right]^{-1}~,\\
p_\alpha^* & \equiv~p_\alpha^{\rm id}(T,\mu_\alpha^*)=\frac{g_\alpha}{6\pi^2}\int_0^\infty \frac{dk\,k^4}{\sqrt{m_\alpha^2+k^2}}\,
\left[\exp\left(\frac{\sqrt{m_\alpha^2+k^2}-\mu_\alpha^*}{T}\right)~-~1\right]^{-1}~,\label{palpha-id}
}
and $m_N \simeq 938$~MeV, $g_N=4$, $m_\alpha \simeq 4m_N-28.3$~MeV, and $g_\alpha=1$.

The system of linear equations (\ref{ni}) can be explicitly solved to yield
\begin{align}
n_N      & = \frac{n^*_N [1+(\tilde b_{\alpha \alpha} - \tilde b_{\alpha N})\,n^*_\alpha]}
{1 + \tilde b_{NN} \, n^*_N + \tilde b_{\alpha \alpha} \, n^*_\alpha + (\tilde b_{NN} \, \tilde b_{\alpha \alpha} -
\tilde b_{N \alpha}\,\tilde b_{\alpha N})\,n^*_N\,n^*_\alpha}~,\\
n_\alpha & = \frac{n^*_\alpha [1+(\tilde b_{N N} - \tilde b_{N \alpha})\,n^*_N]}
{1 + \tilde b_{NN} \, n^*_N + \tilde b_{\alpha \alpha} \, n^*_\alpha + (\tilde b_{NN} \,
\tilde b_{\alpha \alpha} - \tilde b_{N \alpha}\,\tilde b_{\alpha N})\,n^*_N\,n^*_\alpha}~.
\end{align}
Here
\eq{\label{N-id}
n_N^* & \equiv~n_N^{\rm id}(T,\mu_N^*)=\frac{g_N}{2\pi^2}\int_0^\infty dk\,k^2\,
\left[\exp\left(\frac{\sqrt{m_N^2+k^2}-\mu_N^*}{T}\right)~+~1\right]^{-1}~,\\
n_\alpha^* & \equiv~n_\alpha^{\rm id}(T,\mu_\alpha^*)=\frac{g_\alpha}{2\pi^2}\int_0^\infty dk\,k^2\,
\left[\exp\left(\frac{\sqrt{m_\alpha^2+k^2}-\mu_\alpha^*}{T}\right)~-~1\right]^{-1}~.\label{alpha-id}
}
The quantities  $\tilde{b}_{ij}$ are given by Eq~(\ref{bij}) in terms of hard-core radii $r_N$
and $r_\alpha$.

The system of equations (\ref{mu*}) should be solved
with respect to $\mu^*_N$ and $\mu^*_\alpha$ at given $T$ and $\mu$. It reads
\begin{align}
\begin{cases}
\mu_N^*  ~ & ~= ~\mu~ -~ \tilde b_{NN} \, p^*_N ~- ~\tilde b_{N \alpha} \, p^*_\alpha~ + ~2\,a_{NN}\,n_N ~+~(a_{N\alpha} + a_{\alpha N})\,n_\alpha~,\\
 \mu_\alpha^*~ &~ =~ 4\mu ~-~ \tilde b_{\alpha N} \, p^*_N ~-~ \tilde b_{\alpha \alpha} \, p^*_\alpha ~+~2\,a_{\alpha\alpha}\,n_\alpha
~+~(a_{\alpha N} + a_{N\alpha})\,n_N~.
\end{cases}
\end{align}
The system of equations for $\mu^*_N$ and $\mu^*_\alpha$ can be solved numerically.
The system pressure (\ref{p-Nalpha}) and all other thermodynamical
functions can be calculated once the effective chemical potentials $\mu_N^*$ and $\mu_\alpha^*$
are found.
The solution with the largest pressure is taken if multiple solutions appear at a given $\mu$-$T$ pair, in accordance with the Gibbs criterion: This means that only the states corresponding to global thermodynamic equilibrium are considered in the present work.

The following QvdW parameters reproduce the known properties of the nuclear ground state~\cite{Vovchenko:2015vxa}:
\eq{
a_{NN} = 329~\textrm{MeV~fm}^3 \qquad \textrm{and} \qquad \widetilde{b}_{NN} = 3.42~\textrm{fm}^3.
}

For simplicity, attractive vdW interactions involving $\alpha$-particles are neglected, i.e
\eq{
a_{\alpha \alpha} = a_{\alpha N} = a_{N \alpha} =  0,
}
but repulsive EV interactions between
$\alpha$-$\alpha$ pairs and between $\alpha$-$N$ pairs are included,
an effective hard-core radius of $r_\alpha = 1$~fm is assumed.
This entails
\eq{
\widetilde{b}_{\alpha \alpha} = \frac{16 \pi r_\alpha^3}{3}  \simeq 16.76~\textrm{fm}^3.
}
The EV cross terms are calculated according to \eqref{bij}. This gives
\eq{
\widetilde{b}_{\alpha N} \simeq 13.95~\textrm{fm}^3 \qquad \textrm{and} \qquad \widetilde{b}_{N \alpha} \simeq 2.85~\textrm{fm}^3.
}

\begin{table}[h!]
\centering
\setlength{\tabcolsep}{0.5em}
\begin{tabular}{ c | c  c  c  c c }

 & $T_{\rm c}$ (MeV) & $\mu_{\rm c}$ (MeV) & $n_{\rm c}$ (fm$^{-3})$& $p_{\rm c}~ {\rm (MeV\cdot fm^{-3})}$& $X_{\alpha}$ \\
 \hline \hline
 $N$+$\alpha$ mix. & 19.90 & 907.56 & 0.0733 & 0.562 & 0.013 \\
 pure $N$ &  19.68 & 907.67 & 0.0723 & 0.525 & 0
\end{tabular}
\caption{Thermodynamical properties of the mixture of interacting nucleons and $\alpha$-particles at the critical point. 
The results are compared to the case of pure nucleon matter \cite{Vovchenko:2015vxa}. }\label{table:CP-mix}
\end{table}

\begin{figure}[!h]
\centering
\includegraphics[width=0.49\textwidth]{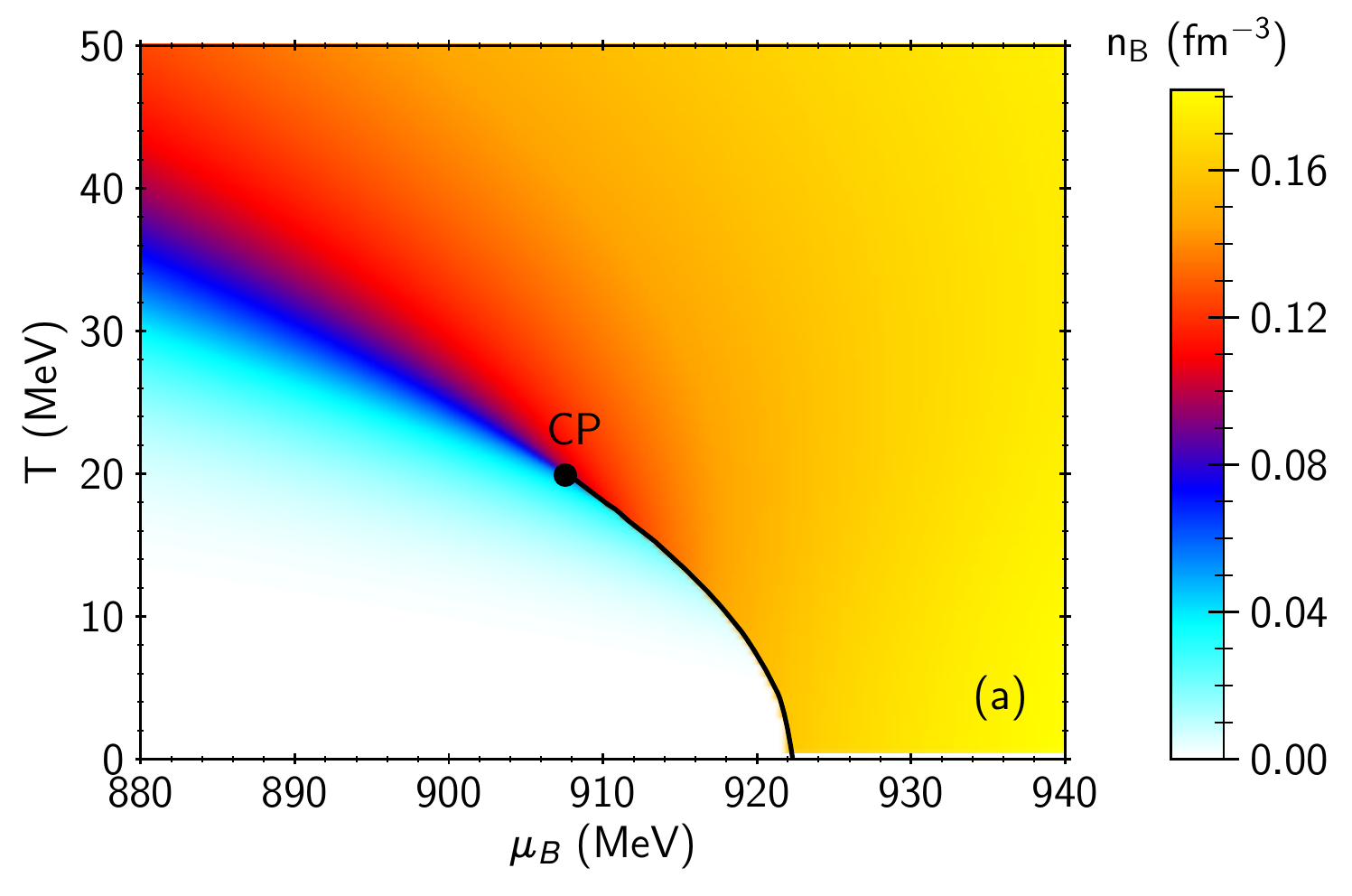}
\includegraphics[width=0.49\textwidth]{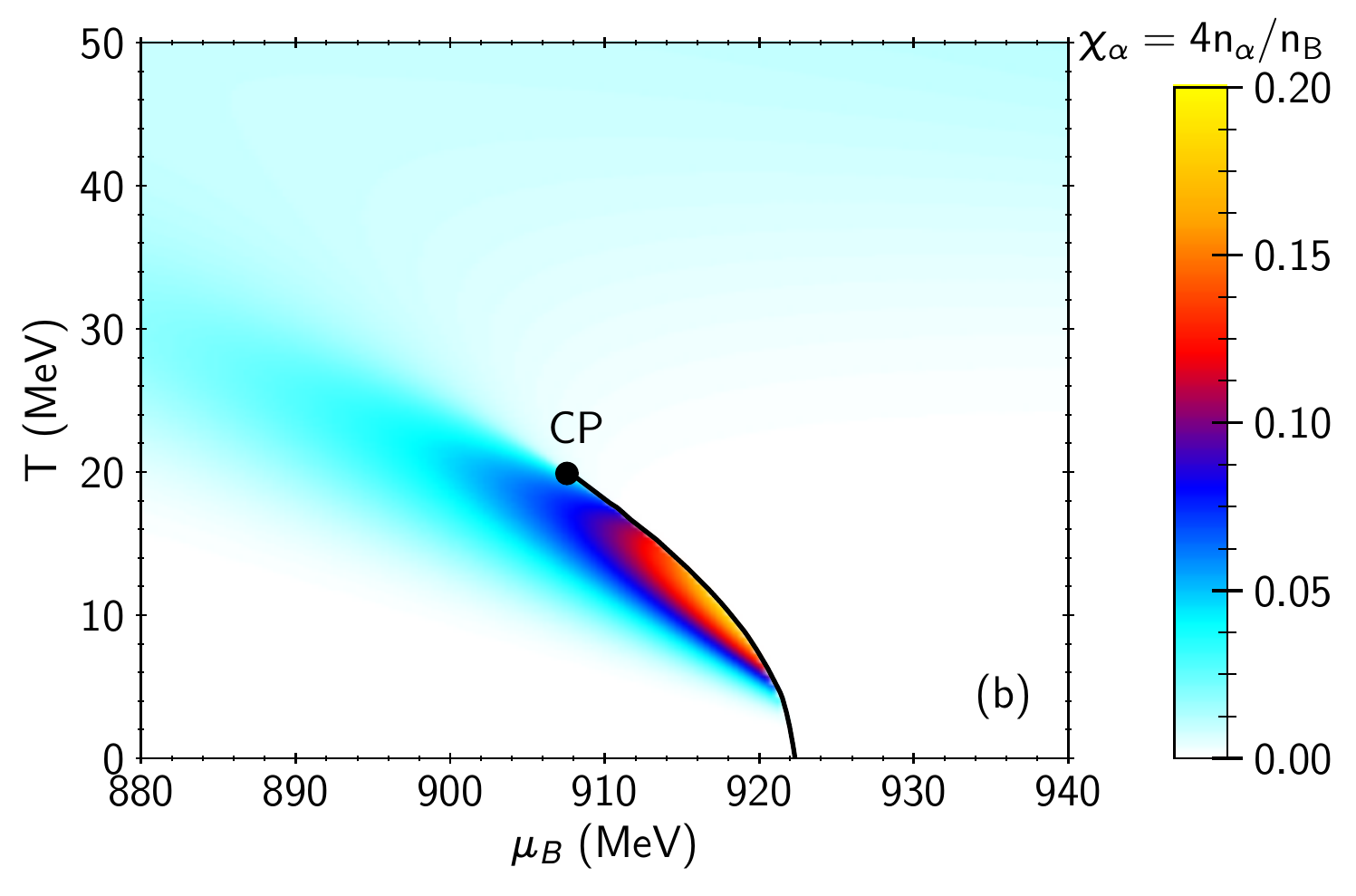}
\caption[]{
Contour plots of the baryonic density $n_B$ (a) and of the 'mass fraction' of the $\alpha$-particles (b) in the $(\mu_B,~T)$ plane.
The solid lines depict the phase transition curve. The full circles correspond to the critical endpoint.
}\label{fig:Phase-Diag}
\end{figure}

The phase diagram for the $N-\alpha$ mixture is shown in Fig.~\ref{fig:Phase-Diag} in the $(\mu,T)$ plane.
The parameters of the critical point (CP)
are presented in Table \ref{table:CP-mix}. 
They are compared to the corresponding results for pure nucleon matter.
The addition of $\alpha$ particles to the model does lead to small changes in the phase diagram of this toy model of nuclear matter.
The phase diagram is very similar to the pure nucleon system~\cite{Vovchenko:2015vxa}.
A slight change in the location of the CP of nuclear matter yields a shift in critical temperature $T_c$ from 19.68~MeV to 19.90~MeV. The critical baryon density $n_c$ increases from 0.0723~fm$^{-3}$ to 0.0733~fm$^{-3}$~(see Table~\ref{table:CP-mix}).
The 'mass fraction' of $\alpha$-particles, $X_\alpha = 4 n_\alpha / (n_N + 4 n_\alpha)$, is approximately, 1.3\% at the CP.

The behavior of the $\alpha$ mass fraction, $X_{\alpha}$, 
is shown
in Fig. \ref{fig:Phase-Diag} (b) 
in the $(\mu, T)$ plane.
$X_\alpha$ jumps from $\simeq 0.2$ to a negligibly small value across the phase transition line at $T = T_c / 2$, 
A noticeable fraction is  present just below the phase transition line.
This jump is caused by the repulsive EV interactions between nucleons and $\alpha$ particles, which suppress the bigger $\alpha$ particles in dense nuclear matter.

\begin{figure}[t]
\centering
\includegraphics[width=0.75\textwidth]{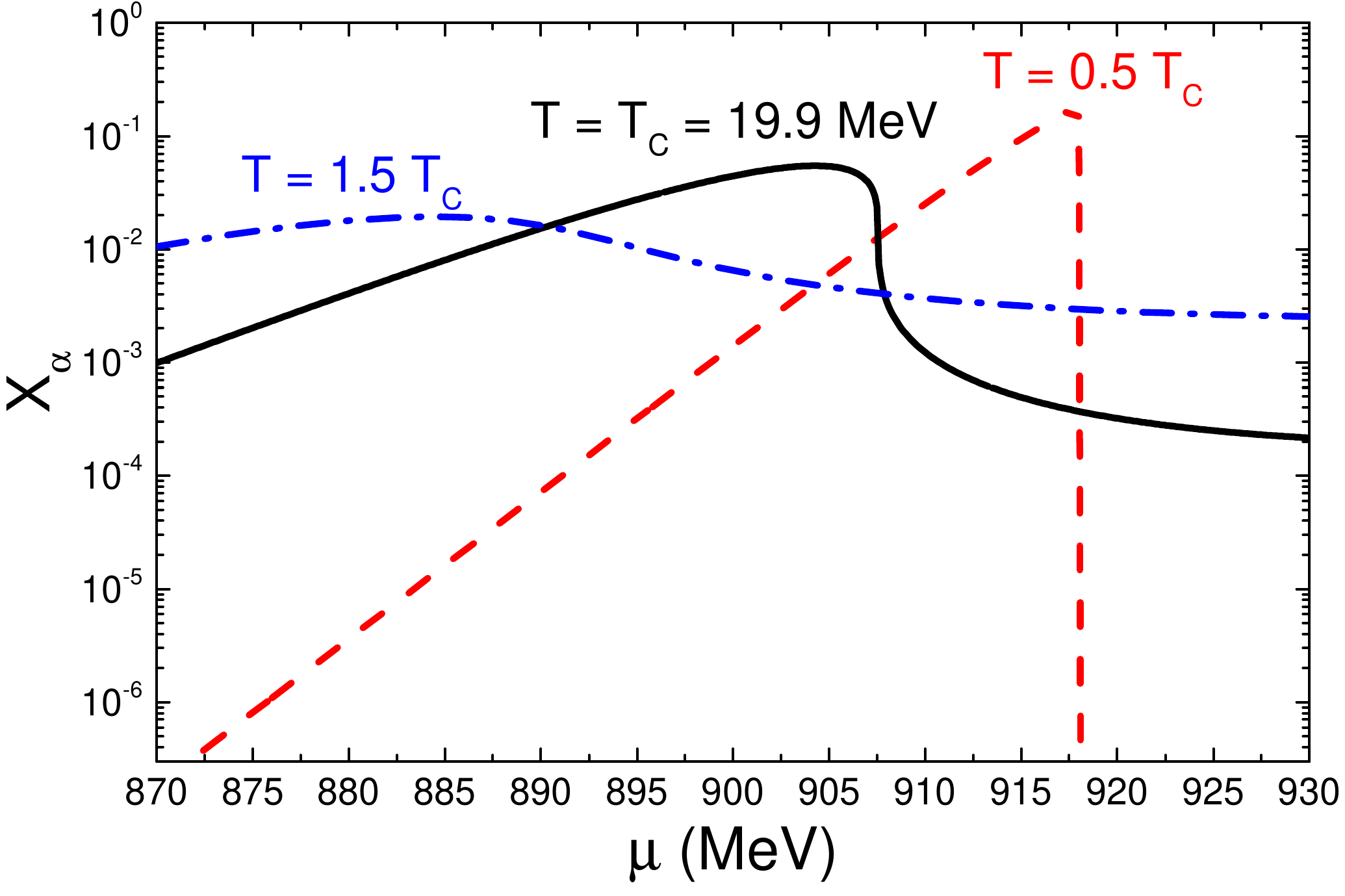}
\caption[]{ 
Dependence of the $\alpha$ mass fraction $X_\alpha$ on the baryon chemical potential $\mu$ for three different isotherms.
The dashed red line corresponds to $T = T_c \,/ \, 2 \simeq 10$~MeV, half the critical temperature. The solid black line corresponds to the critical isotherm $T = T_c \simeq 19.89$~MeV. The dash-dotted blue line corresponds to $T = 3  \, T_c \, / \, 2 \simeq 30$~MeV, above the critical temperature.
}\label{fig:N-alpha-isotherms}
\end{figure}

The dependence of $X_\alpha$ on the baryochemical potential $\mu$ is shown in Fig.~\ref{fig:N-alpha-isotherms} for three different isotherms.
The dashed red line depicts the isotherm $T = 10$~MeV, at half the critical temperature.
$X_\alpha$ increases smoothly with chemical potential until reaching the liquid-gas coexistence curve. 
Higher values of $\mu$ correspond to a dense liquid composed mainly of nucleons. 
The $\alpha$ mass fraction drops to negligibly small values in the liquid phase.
As mentioned above, only the globally stable thermodynamic states are considered.
The metastable parts of the $T = 10$~MeV isotherm~(which also exist) are not shown in Fig.~\ref{fig:N-alpha-isotherms}.

The behavior of $X_\alpha$ at the critical isotherm $T = T_c$ (solid black line in Fig.~\ref{fig:N-alpha-isotherms}) is qualitatively similar to the previous one. 
$X_\alpha$ exhibits a rapid, but continuous drop at CP and in the vicinity of the CP along the critical isotherm. This correlates with a rapid increase of the baryon density with $\mu$ across the critical isotherm near the CP.

Finally, the isotherm $T = 3 \, T_c \, / \, 2 \simeq 30$~MeV in the
crossover region is shown in Fig.~\ref{fig:N-alpha-isotherms} as the dash-dotted blue line.
The $\mu$-dependence of $X_\alpha$ shows a broad maximum at this isotherm.
This bump corresponds to the crossover region of the phase diagram~(Fig.~\ref{fig:Phase-Diag}), where a 
smooth increase of the baryon density with $\mu$ takes place.

The example presented in this subsection shows that the repulsive vdW interactions cause the cluster dissolution at high baryonic densities.
This conclusion is not new: the EV interactions had previously been used in relativistic mean field models of nuclear matter~\cite{Lattimer:1991nc,Shen:1998by,Typel:2016srf}.
More realistic studies of nuclear matter must also take into account the attractive interactions involving $\alpha$ particles.

Presented studies can be extended.
For instance, only globally stable thermodynamic states of the $N$-$\alpha$ mixture were considered in this work: the Gibbs criterion in the GCE was applied. A more complete picture can be obtained by additionally analyzing metastable and unstable states.
This work assumes that $\alpha$ particles do not form Bose-Einstein condensates.
The effects of Bose-Einstein condensates of $\alpha$ particles could play a significant role~\cite{Satarov:2017jtu}, especially at low temperatures.
Other nuclear clusters, such as $d$, $t$, etc., should be included in the description of nuclear matter as well~\cite{Hahn:1986mb}.

\subsection{Flavor-dependent vdW interactions in HRG and lattice data at $\mu_B = 0$}

Flavor-dependent vdW interactions in the HRG model can be considered in the context of the lattice QCD data.
The strong influence of the baryon-baryon vdW interactions 
was demonstrated for
observables 
accessible with lattice QCD 
at zero chemical potential in the crossover region~\cite{Vovchenko:2016rkn}, using
identical vdW interactions between only (anti)baryon pairs.
The vdW terms between all other hadron pairs were neglected in~\cite{Vovchenko:2016rkn}.
This simplest scenario can be modeled essentially with a single-component quantum statistical vdW model.
Evidently, different baryon pairs may have different vdW parameters.
Heavier strange hadrons may have different parameters compared to non-strange ones, e.g. a smaller size.
A thermal analysis of the hadron yield data~\cite{Alba:2016hwx} suggests this possibility.

Presently vdW interactions are included only for (anti)baryon pairs, similarly to Ref.~\cite{Vovchenko:2016rkn}.
We extend the QvdW-HRG model\footnote{In the notation of Ref.~\cite{Vovchenko:2016rkn} it is referred to as the VDW-HRG model.} of Ref.~\cite{Vovchenko:2016rkn} by considering grossly different vdW parameters for non-strange and strange baryons.
The vdW parameters which reproduce properties of the nuclear ground state~\cite{Vovchenko:2015vxa} are employed for all pairs of non-strange baryons, i.e. 
$a_{\rm NS} = 329~\textrm{MeV~fm}^3$ 
and 
$\widetilde{b}_{\rm NS} = 3.42~\textrm{fm}^3$.
The effective hard-core radius of strange baryons is assumed to be half that of non-strange baryons. 
Hence, the EV parameter $b_{\rm S}$ of a strange baryon is 8 times smaller than the EV parameter $b_{\rm NS}$ of non-strange baryon.
The attractive vdW parameters $a_{\rm NS}$ here are assumed to be also a factor 8 smaller then the non-strange ones. Thus,
$a_{\rm S} = a_{\rm NS} / 8 \simeq 41~\textrm{MeV~fm}^3$
and
$\widetilde{b}_{\rm S} = \widetilde{b}_{\rm NS}/8 \simeq 0.43~\textrm{fm}^3$.

A large, factor 8, difference between vdW parameters for strange and non-strange baryons 
in the present toy model
illustrates
the multi-component vdW formalism. 
Obviously, a different scenario with revised values of these parameters may be considered as well.   
The vdW interactions between strange and non-strange baryons are characterized by the corresponding cross term coefficients.
The repulsive cross term coefficients $\widetilde{b}_{ij}$ are calculated according to Eq.~\eqref{bij}.
The attractive cross term coefficients $a_{ij}$ are calculated as a geometric mean, i.e.
\eq{
a_{ij} = \sqrt{a_i \, a_j}.
}
This particular mixing rule is motivated by its common use in chemistry~\cite{ChemMix1,ChemMix2}.

\begin{figure}[t]
\begin{center}
\includegraphics[width=0.49\textwidth]{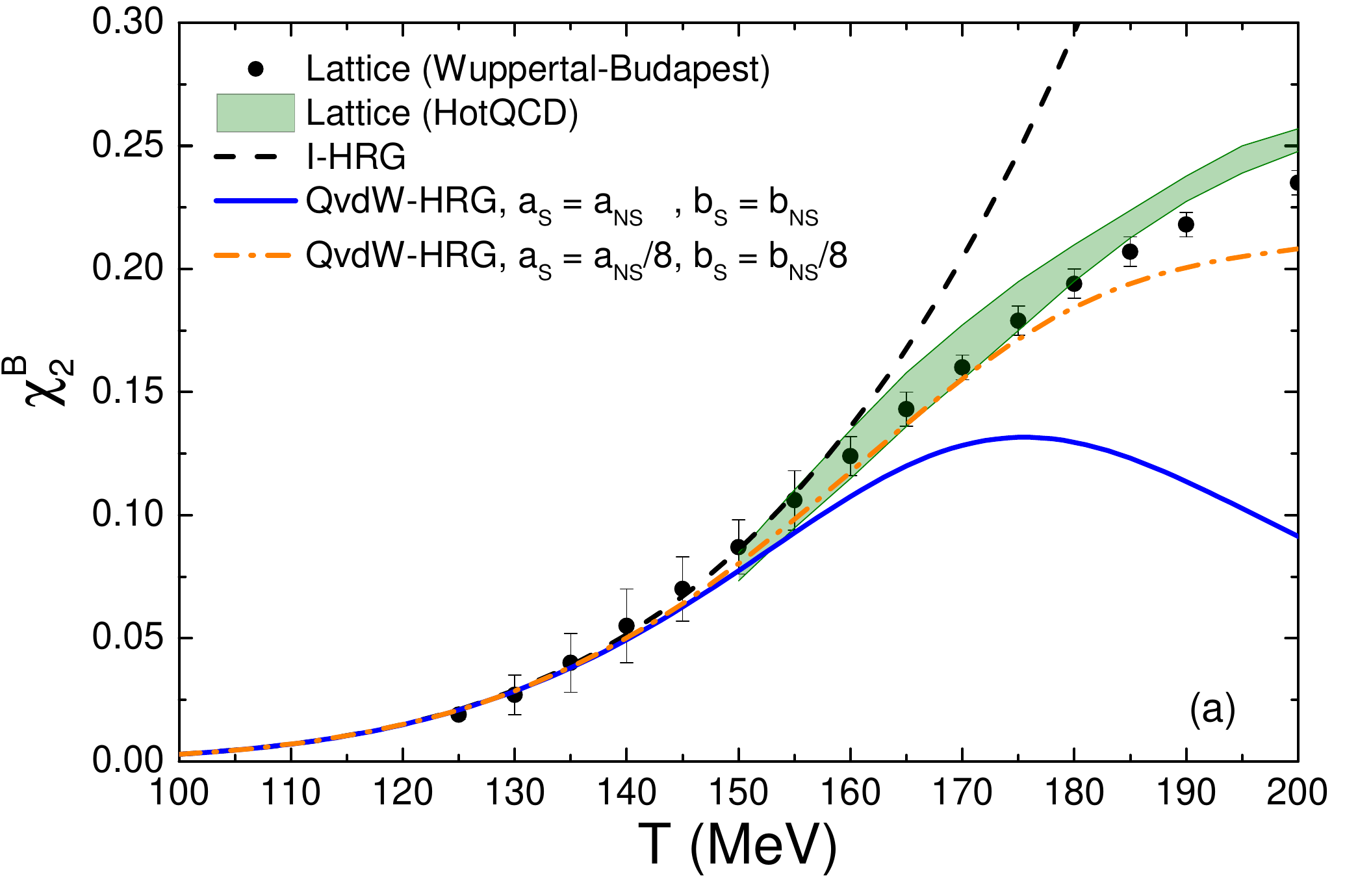}
\includegraphics[width=0.49\textwidth]{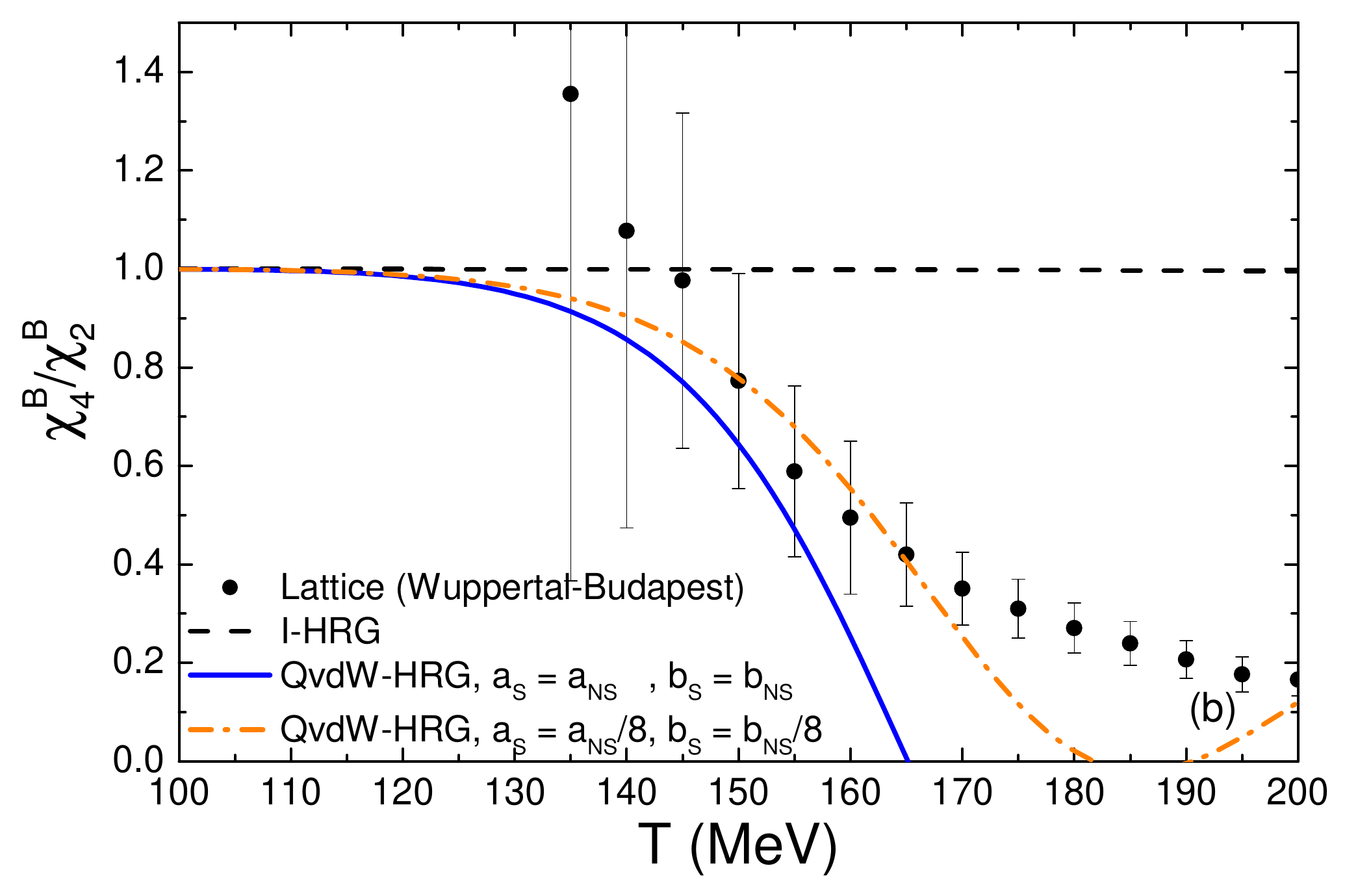}
\caption{The temperature dependence of
(a) $\chi_2^B$ and (b) $\chi_4^B / \chi_2^B$ net baryon number susceptibilities, as calculated within the I-HRG model (dashed black lines), the QvdW-HRG model with $a_{\rm S} = a_{\rm NS}$ and $\widetilde{b}_{\rm S} = \widetilde{b}_{\rm NS}$ (solid blue lines),  and the QvdW-HRG model with $a_{\rm S} = a_{\rm NS} / 8$ and $\widetilde{b}_{\rm S} = \widetilde{b}_{\rm NS} / 8$ (dash-dotted orange lines), at zero chemical potential.
The lattice QCD results of the Wuppertal-Budapest~\cite{Borsanyi:2011sw,Bellwied:2015lba} and HotQCD~\cite{Bazavov:2012jq} collaborations are shown, respectively, by symbols and green bands.
}
\label{fig:SNS:chiB}
\end{center}
\end{figure}

The temperature dependences of the net baryon susceptibilities $\chi_2^B$ and $\chi_4^B / \chi_2^B$ are calculated 
for the different models:
within the I-HRG model, the QvdW-HRG model with $a_{\rm S} = a_{\rm NS}$ and $\widetilde{b}_{\rm S} = \widetilde{b}_{\rm NS}$, and the QvdW-HRG model with $a_{\rm S} = a_{\rm NS} / 8$ and $\widetilde{b}_{\rm S} = \widetilde{b}_{\rm NS} / 8$, at $\mu = 0$. These dependences are compared in Fig.~\ref{fig:SNS:chiB} to the lattice QCD data of the Wuppertal-Budapest~\cite{Borsanyi:2011sw,Bellwied:2015lba} and HotQCD~\cite{Bazavov:2012jq} collaborations.
The scenario with weaker vdW interactions involving strange baryons improves the agreement with the lattice data.
This is primarily caused by the overall decrease of the effects of the repulsive EV interactions.

\begin{figure}[t]
\begin{center}
\includegraphics[width=0.49\textwidth]{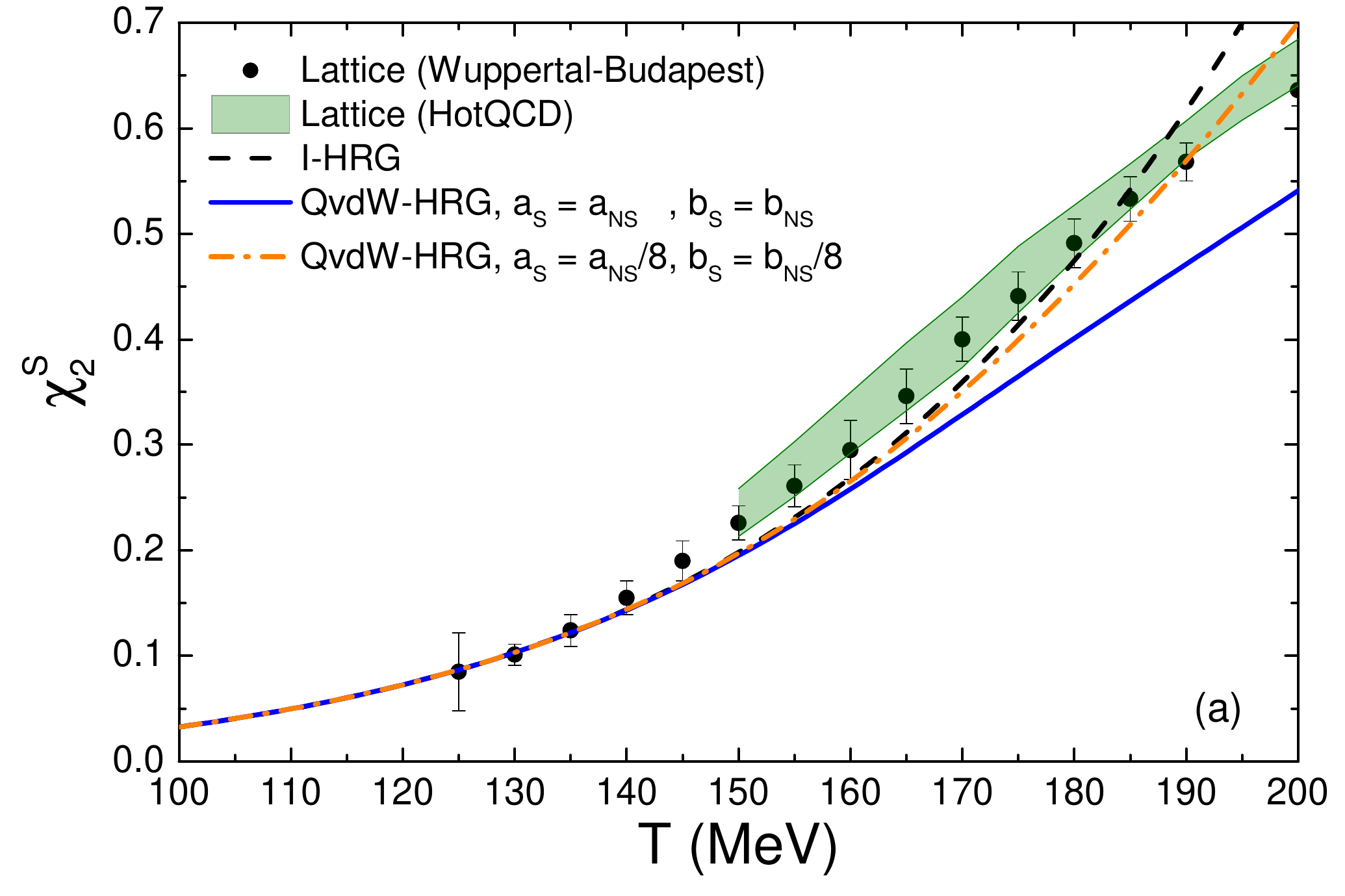}
\includegraphics[width=0.49\textwidth]{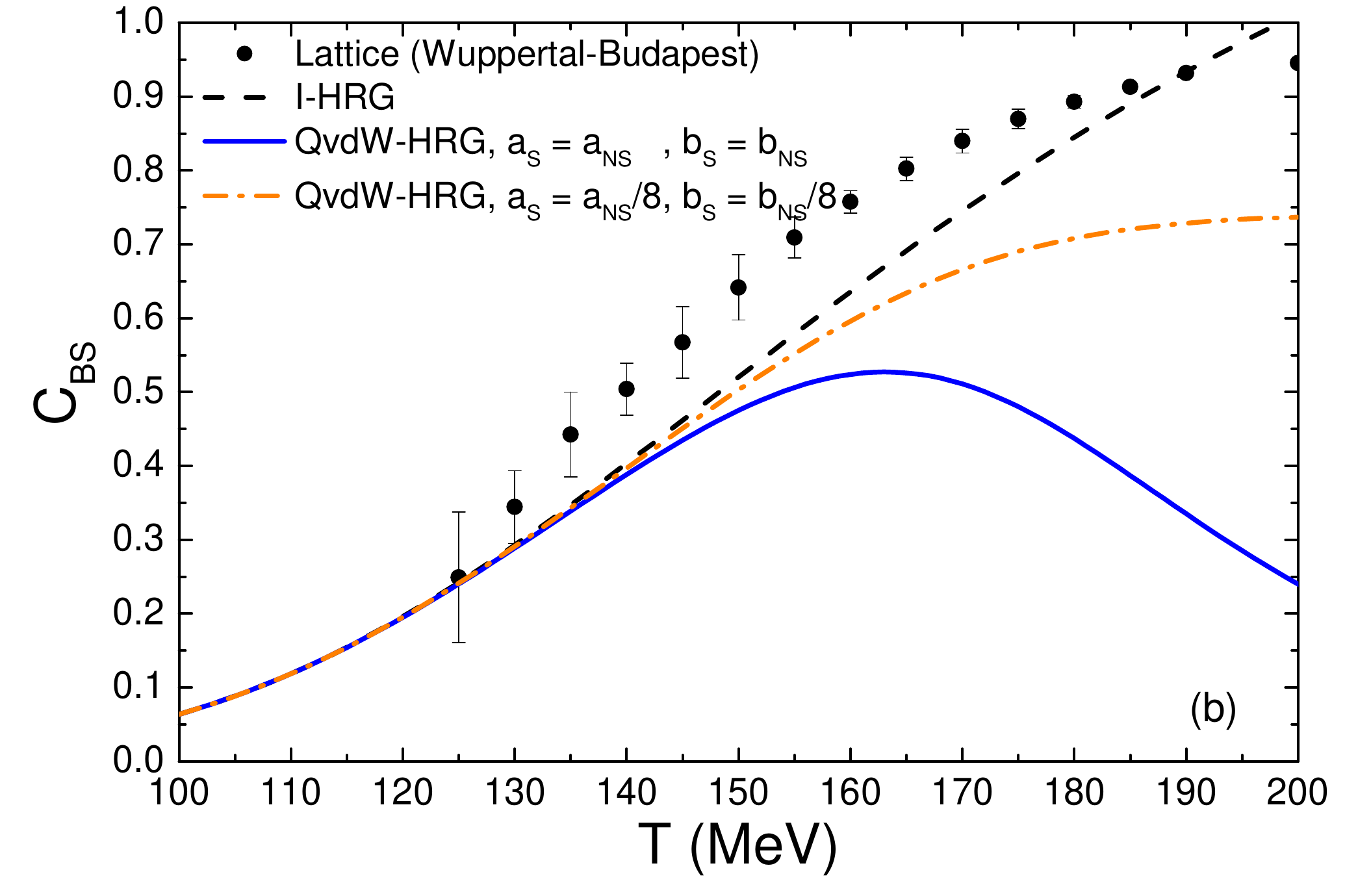}
\caption{The temperature dependence of
(a) net-strangeness susceptibility $\chi_2^S$ and (b) baryon-strangeness correlator ratio $C_{BS} = -3 \, \chi_{11}^{BS} / \chi_2^S$, as calculated within the I-HRG model (dashed black lines), the QvdW-HRG model with $a_{\rm S} = a_{\rm NS}$ and $\widetilde{b}_{\rm S} = \widetilde{b}_{\rm NS}$ (solid blue lines),  and the QvdW-HRG model with $a_{\rm S} = a_{\rm NS} / 8$ and $\widetilde{b}_{\rm S} = \widetilde{b}_{\rm NS} / 8$ (dash-dotted orange lines), at zero chemical potential.
The lattice QCD results of the Wuppertal-Budapest~\cite{Borsanyi:2011sw} and HotQCD~\cite{Bazavov:2012jq} collaborations are shown, respectively, by symbols and green bands.
}
\label{fig:SNS:chiS}
\end{center}
\end{figure}

The strangeness observables are sensitive to vdW interactions of strange baryons. The net strangeness susceptibility $\chi_2^S$ and the baryon-strangeness correlator ratio $C_{BS} = -3 \, \chi_{11}^{BS} / \chi_2^S$ are used to demonstrate this. 
The latter observable, suggested in Ref.~\cite{Koch:2005vg}, is particularly sensitive and therefore is a useful diagnostic tool for QCD matter.
The calculations for these two observables are shown in Fig.~\ref{fig:SNS:chiS}.
They are compared with the corresponding lattice results of the Wuppertal-Budapest~\cite{Borsanyi:2011sw} and HotQCD~\cite{Bazavov:2012jq} collaborations.

The standard QvdW-HRG model (identical vdW parameters for both, the non-strange and strange baryons) does not improve the agreement with the lattice data as compared to the I-HRG model for these observables. 
In fact, the agreement becomes significantly worse at high temperatures.
In the scenario with the smaller vdW interactions for strange baryons, on the other hand, the existing agreement of the I-HRG model with the lattice QCD data for $\chi_2^S$ is preserved, see Fig.~\ref{fig:SNS:chiS}a.

None of the considered scenarios describes the lattice data for the $C_{BS}$~(Fig.~\ref{fig:SNS:chiS}b):
The lattice data are underestimated by all three models.
The two QvdW-HRG models do show a characteristic inflection point in the temperature dependence of $C_{BS}$. This point seems to be present in the lattice QCD data as well.
Reducing the vdW interactions of strange baryons does result in an improved agreement with the lattice QCD data.
Hitherto undiscovered strange hadrons may be the
source of the underestimation of $C_{BS}$ in HRG models.
These states have been predicted by the quark model~\cite{Capstick:1986bm,Ebert:2009ub}
and by the lattice QCD spectrum calculations~\cite{Edwards:2012fx}.
The agreement between the lattice QCD data and the I-HRG model for $C_{BS}$ is improved if these extra states are included into the I-HRG model~\cite{Bazavov:2014xya}.
A similar effect is expected for all QvdW-HRG based models.

\section{Summary} \label{Sum}

This paper presented a generalization of the van der Waals equation of state for a multi-component system.
The formalism takes into account both the repulsive and the attractive interactions between particles of different species.
It allows to specify the parameters characterizing repulsive and attractive forces between each pair of species independently. 
The quantum statistical effects, absent in the classical van der Waals equation, are introduced into the free energy of the multi-component van der Waals model.
The grand canonical ensemble formulation of the model is also presented.

Both, the grand canonical formulation and the implementation of the quantum statistics in the multi-component van der Waals model, have been done in the present paper for the first time.
These extensions are useful for many physical applications.
The quantum statistical formulation
is helpful for describing 
the asymmetric nuclear matter, 
which consists of an unequal number of interacting protons and neutrons. The formulation also allows treatment of the light nuclei in dense nuclear matter.
The repulsive and attractive van der Waals interactions seem to be crucially important for baryons and antibaryons in hot hadronic matter.
Such an application of the multi-component van der Waals model is discussed
in the present paper in the context of the strangeness dependent interactions. 
The formalism developed in this paper should be also useful
for studies of the multi-component systems of interacting atoms and molecules 
at small temperatures, where the quantum statistical effects are important.

\vspace{0.5cm}
\section*{Acknowledgements}
This work was supported by the Helmholtz International Center for FAIR within the LOEWE program of the State of Hesse.
V.V. acknowledges the support from HGS-HIRe for FAIR. A.M. is thankful for the
support from Norwegian Centre for International Cooperation in Education, Grant No. CPEA-LT-2016/10094.
The work of M.I.G. is supported by the
Goal-Oriented Program of  the National Academy of Sciences of Ukraine and
the European Organization for Nuclear Research (CERN), Grant CO-1-3-2016 and
by the Program of Fundamental Research of the Department of Physics
and Astronomy of National Academy of Sciences of Ukraine.
L.M.S. acknowledges the support from the Frankfurt Institute for Advanced Studies.
H.St. appreciates the support from J.M.~Eisenberg Laureatus chair at the Goethe University Frankfurt.

\end{document}